\documentstyle[12pt]{article}

\begin{document}

\begin{center}
\Large{{\bf Tomographic and statistical properties of superposition
states for two-mode systems}}
\end{center}
\begin{center}
Sergey V. Kuznetsov$^1$, Aleksander V. Kyusev$^1$  
and Olga V. Man'ko$^2$\\
$1$ -Physics Department, M.V. Lomonosov Moscow State University,
Vorobievy Gori, Moscow 119899, Russia \\
$2$ - Lebedev Physical Institute, Leninsky Prospect 53, 117924 Moscow, Russian Federation\\
\end{center}
 E-mail: mankoov@lebedev.ru \\

\begin{abstract}
The two-mode even and odd coherent states and two-mode squeezed correlated
state are discussed.
Photon distribution functions, means, dispersions, Fano factor for even 
and odd coherent states and squeeezed correlated state are calculated. 
The photon distribution function for two-mode squeezed correlated state is
obtained. The tomograms of two-mode even and odd coherent states and
squeezed correlated states are 
investigated within the framework of a symplectic tomography scheme.
It is shown that the phenomenon of entanglement appears
in the system. Two different measures of entanglement are employed.
\end{abstract}
\vspace{3mm}

\section{Introduction}
The odd and even coherent states were introduced in~\cite{[1],[2]}.
In this works, the name "even and odd" coherent states has been given
to even and odd superpositions of two Gaussian packets describing
coherent states. In~\cite{[3]} these states were discussed as a
subclass of some generic set of nonclassical states. A scheme of
generation of even and odd coherent states of a trapped ion have been
proposed in~\cite{[4]}. This scheme gives the possibility of
studying quantum interference phenomena with essentially higher stability
than realization of the even and odd coherent states in quantum
optics~\cite{[5]}. The importance of the even and odd coherent states
of the electromagnetic fields is also related to the possibilities of
reducing the noise influence on the signal in the process of quantum-state
signal transmission used in optical communication \cite{[6]}. Also, the
even and odd coherent states might be used as alternatives to squeezed
states of light to improve the sensitivity of interferomagnetic
gravitational wave detectors~\cite{[7]}. For large amplitudes of the
partners of the superposition of two coherent states, these states of
light and their slight modifications were interpreted as "Schr\"odinger
cat states" in~\cite{[8]} where their generation due to propagation of
initially coherent light in Kerr medium was suggested. One can use
generalized correlated states~\cite{[9]} as the partners of the
superposition to take into account the influence of mode quadratures
correlations on the nonclassical properties of light. 
Last years quantum statistical properties of odd and coherent states
have been the subject of intensive
experimental and theoretical investigations and still remain the
subject of study~\cite{[10]}-\cite{[19]}.   
 
Generic systems with quadratic Hamiltonians 
(multivariable parametric oscillators) have statistical properties
of their Fock states (number eigenstates) described by multivariable
Hermite polynomials~\cite{[20]}-\cite{[23]}. 
General formulas for matrix elements of the Gaussian density operator for a 
multimode oscillator in Fock basis were calculated explicitly in~\cite{[20]}. 
The photon distribution for an $N$-mode mixed state of light described by a Wigner 
function of the generic Gaussian form was calculated explicitly in terms of Hermite 
polynomials of $2N$ variables in~\cite{[21]} and the parameters of the photon 
distribution function were determined through the dispersion matrix and mean values 
of quadrature components of the light. The photon distribution for the two-mode 
squeezed vacuum was investigated in~\cite{[22]} where 
its dependence on four parameters (two squeezing coefficients, relative phase 
between the two oscillators, and their spatial orientation) was shown. 
In~\cite{[23]} the case of generic two-mode squeezed coherent states was 
considered, and the photon distribution function for the states was expressed through 
both four-variable and two-variable Hermite polynomials dependent on two squeezing 
coefficients, the relative phase between the two oscillators, their spatial
orientation, and four-dimensional shift in the phase space of the
electromagnetic-field oscillator. In~\cite{[24]} the multimode even and
odd coherent states were considered. The explicit formulae for the photon
distribution, Wigner function were derived. In~\cite{[25]} the multimode
Schr\"odinger cat state were constructed for polymode parametric oscillators of
the electromagnetic field. The evolution of the photon distribution function
was evaluated explicitly, the distribution function were expressed in terms of
multivariable Hermite polynomials.

In two-mode odd and even coherent states and two-mode squeezed correlated
states, one has the interaction of the photon modes, which creates
different correlations of the mode properties. Such correlations can be
considered as phenomenon of the mode entanglement. The entanglement of
subsystems of composite systems is important ingredient of quantum
information processing and quantum computing. 
It is worthy to clarify the properties of
entanglement between the modes in the states under study. We investigate
two measures of entanglement introduced in~\cite{[26],[27]} for two-mode
squeezed correlated state and two-mode even and odd coherent states.

Recently the tomographic method of state reconstructing was introduced
for generic quantum systems in \cite{[28]}-\cite{[30]}. 
One measures the state tomogram which is the standard probability
distribution function. The reconstruction formula gives the possibility
to obtain the density matrix of the state. For optical tomography, this
method was realized for measuring photon states (see, for example,
\cite{[31],[32]}). The tomographic approach was used to present the
new formulation of quantum mechanics in which quantum states are
associated with standard probability
distributions~\cite{[33]}-\cite{[35]}. The sympectic tomograms of
Schr\"odinger cat states of a trapped ion were investigated
in~\cite{[36],[37]}. 
The two-mode even and odd coherent states and two-mode squeezed
correlated states can be associated with the tomograms which, being the
standard probability distributions, contain the same information on the
states as the wave function or density matrix does.

The aim of the paper is to consider two-mode even and odd coherent
states and two-mode squeezed correlated states, to study their photon
distribution functions, to consider the phenomenon of
entanglement between the modes in both states and to obtain in explicit form
their tomograms within the framework of symplectic tomography scheme.

The paper is organized as follows. In Sec.~1 we discuss the wave function,
Wigner function and dispersions of quadratures in pure two-mode squeezed
correlated states.
In Sec.2, we concentrate on the generic case of the 
photon statistics of a pure two-mode squeezed coherent state and obtain
for it two explicit formulae through Hermite polynomials of four variables
and through Hermite polynomials of two variables.
In Sec.~3 we discuss the averaging of photon distribution function of
two-mode squeezed correlated states over one mode and obtain the probability
to have $n$ photon in other mode. In Sec.~4 we investigate the phenomenon
of the entanglement in the two-mode squeezed correlated states.
We present in Sec. 5 the exact formulae for tomogram of two-mode squeezed
correlated states within the framework of symplectic tomography schemes.
In Sec.~6 we discuss the wave function, Wigner function, photon number
probability distribution function and dispersions of quadratures and photon
numbers in the modes, concentrate on the phenomenon of entanglement and
obtain the tomogram within the framework of symplectic tomography for
two-mode even and odd coherent states.

\section{Two-mode squeezed correlated states}
In this section, we discuss the wave function, Wigner function
and dispersion of quadratures in two-mode squeezed correlated states.
The wave function $\Psi_{\rm sq}(x_1,x_2)$ that described 
the generic form of the photon statistics of a pure two-mode
squeezed coherent state was found in~\cite{[23]} and is of the form 
\begin{equation}   
\Psi_{\rm sq}(x_1,x_2) = {\cal N} \, \exp\left(-A x_1^2- B x_2^2+2C x_1x_2+D x_1   +E x_2\right),
\label{eq.2.2}
\end{equation} 
with the normalization constant 
\begin{equation}   
{\cal N} = \sqrt{\frac{2}{\pi}}\left(A_1 B_1 - C_1^2\right)^{1/4}   
\exp \left[ - \frac{1}{4 (A_1 B_1-C_1^2)}    
\left(B_1 D_1 D + A_1 E_1 E + C_1 \left[E_1 D + D_1 E\right]   \right)
\right].
\label{eq.norm}
\end{equation}
The wave function is a shifted Gaussian function described by the
five complex numbers
\begin{eqnarray*}  
 A=A_1+i A_2,\quad   B=B_1+i B_2,\quad C=C_1+i C_2,\quad  D=D_1+i D_2,\quad    
E=E_1+i E_2.
\end{eqnarray*} 
The Wigner function of squeezed correlated states can be represented
in the form  
\begin{equation}   
W(\vec{Q})=(2\pi)^{-2}(\det \sigma)^{-1/2} \exp\left[-{1 \over  2}
\Big [\Big (\vec{Q}-\langle\vec{Q}\rangle\Big)\sigma^{-1}   
\Big(\vec{Q}-\langle\vec{Q}\rangle\Big)\Big ]\right],
\label{wigmean}
\end{equation} 
where $\sigma$ is the quadrature dispersion matrix and the 
four-dimensional vector $\vec{Q}$ has the components $(p_1,p_2,x_1,x_2)$. 
The four-dimensional vector $\langle\vec{Q}\rangle$ has the quadratures 
means as its components:
\begin{eqnarray}   
&&\langle p_1 \rangle  =  \frac{E_1 ( A_1 C_2 - A_2 C_1 ) -   
D_1 (A_2 B_1 - C_1 C_2 ) + D_2 (A_1 B_1 - C_1^2)}   {A_1 B_1 - C_1^2}\,,   
\nonumber \\   
&&\langle p_2 \rangle  =  \frac{D_1 ( B_1 C_2 - B_2 C_1 ) -   E_1 (A_1 B_2 - 
C_1 C_2 ) + E_2 (A_1 B_1 - C_1^2)}   {A_1 B_1 - C_1^2}\,,   
\nonumber \\
&&
\langle x_1 \rangle  =  \frac{B_1 D_1 + C_1 E_1}{2 \, (A_1 B_1 - C_1^2)},\quad    
\langle x_2 \rangle = \frac{A_1 E_1 + C_1 D_1}{2 \, (A_1 B_1 - C_1^2)}\,.
\label{means}
\end{eqnarray}
The dispersion matrix $\sigma$ with the matrix elements: 
\begin{eqnarray}   
&&\sigma_{x_1,x_1} = \frac{B_1}{4 \, (A_1 B_1 - C_1^2)},\quad   
\sigma_{x_1,x_2}  = \frac{C_1}{4 \, (A_1 B_1 - C_1^2)},\quad
\sigma_{x_1,p_1} =  \frac{C_1 C_2 - B_1 A_2 } {2 \, (A_1 B_1 - C_1^2)}\,,   
\nonumber \\   
&&\sigma_{x_1,p_2}  =  \frac{C_2 B_1 - B_2 C_1 } {2 \, (A_1 B_1 - C_1^2)},\quad    
\sigma_{x_2,x_2}  =  \frac{A_1}{4 \, (A_1 B_1 - C_1^2)},\quad  
\sigma_{x_2,p_1}  =  \frac{C_2 A_1 - A_2 C_1 } {2 \, (A_1 B_1 - C_1^2)},\quad
\nonumber \\   
&&\sigma_{x_2,p_2} =  \frac{C_1 C_2 - A_1 B_2 } {2 \, (A_1 B_1 - C_1^2)},\quad    
\sigma_{p_1,p_1}  =  \frac{A_1^2 B_1 - A_1 C_1^2 + A_1 C_2^2 +    A_2^2 B_1 - 2 \, 
A_2 C_1 C_2}{A_1 B_1 - C_1^2}\,,   \nonumber \\   
&&\sigma_{p_1,p_2} =  \frac{- A_1 B_1 C_1 + A_2 B_2 C_1 + C_1^3     + 
C_1 C_2^2 - A_1 B_2 C_2 - A_2 B_1 C_2}{A_1 B_1 - C_1^2}\,,   \nonumber\\
&&\sigma_{p_2,p_2} =  \frac{A_1 B_1^2 - B_1 C_1^2 + B_1 C_2^2 +    B_2^2 A_1 - 2 \,
 B_2 C_1 C_2}{A_1 B_1 - C_1^2}\,.
\label{eq.sigma}
\end{eqnarray} 
The matrix elements of the dispersion matrix $M$ can also be calculated by using the 
wave function~(\ref{eq.2.2}). The determinant of the dispersion matrix $M$ can be 
checked to be equal to $1/16$, which means that our squeezed state~(\ref{eq.2.2}) 
minimizes the generalized Schr\"odinger uncertainty relation~\cite{[38]} 
\[   
\mbox{det}\, M \geq \frac{1}{16}\,.
\] 
In this sense, the wave function $\Psi_{\rm sq} $ is called a minimum
uncertainty state.

\section{Photon Distribution Function for two-mode squeezed correlated
states}
In this section, we concentrate on the generic case of the 
photon statistics of a pure two-mode squeezed coherent state. 
In order to calculate the distribution function
$W\left (n_1,\,n_2\right )$, we construct the two-mode squeezed coherent 
wave function~(\ref{eq.2.2})
with the probability amplitude of the photon-energy states
\begin{equation}
   \Psi_{n_1,\,n_2}(x_1,\,x_2)=\prod _{i=1}^2 \pi^{-1/4}\,
\frac{1}{\sqrt{2^{n_i}\,n_i!}}\,e^{-x_i^2/2}\,
   H_{n_i}(x_i)\,,
\end{equation}
where $H_n$ denotes the $n$th Hermite polynomial, and arrive at
\begin{eqnarray}
&&   P(n_1,n_2)=  \frac{\vert {\cal N} \vert^2}{\pi \,2^{n_1+n_2}\,
   n_1!\,n_2!}\,
   \Bigg| \int_{-\infty}^{ \infty} \exp\left[-\left(A+\frac {1}{2}\right)
x_1^2-\left(B+\frac {1}{2}\right)x_2^2\Bigg.\right.,\nonumber\\
&&\left.+2C x_1x_2 +D x_1 +E x_2\right] 
   H_{n_1}(x_1)H_{n_2}(x_2)\, dx_1 \,dx_2 \,\Bigg| ^2.
\end{eqnarray}
We calculate this integral and obtain
\begin{equation}
   P(n_1,n_2)=W(0,0)\, \frac{1}{n_1! \, n_2!}\,
   \left| H_{n_1,n_2}^{\{R\}}(y_1,y_2) \right| ^2,
\label{eq.4.1}\end{equation}
with the probability to have no photon in any of the modes
\begin{eqnarray}
   P(0,0) & =& \frac{2 \, \sqrt{A_1 B_1 - C_1^2}}{|(A+1/2)(B+1/2)-C^2|}
  \, \exp \left[ - \frac{A_1 E_1^2 + B_1 D_1^2 + 2 C_1 D_1 E_1 }
   {2 \, (A_1 B_1 - C_1^2)} \right]
   \nonumber \\ & &
   \times \left| \exp \left[ \frac{ (A+1/2)E^2 + (B+1/2) D^2 - 2CDE }
   { 2 \, ((A+1/2)(B+1/2) - C^2) } \right] \right|  .  
\end{eqnarray}
In Eq. (\ref{eq.4.1}), $H_{n_1,n_2}^{\{R\}}(y_1,y_2)$ denotes the
two-dimensional Hermite polynomial. 
We use the generating function for the Hermite polynomials 
\begin{equation}
   \exp \left (2 t x -x^2\right )=\sum_{n=0}^{\infty}\frac{x^n}{n!}\,H_n(t)\,,
\end{equation}
The matrix $R$ and the vector
$\vec{ y}$ in Eq.~(\ref{eq.4.1}) are given by the relations
\begin{equation}
   R = \frac{1}{ (A+1/2)(B+1/2)-C^2 }
   \pmatrix{(A-1/2)(B+1/2)-C^2 & -C \cr 
   -C & (A+1/2)(B-1/2)-C^2 }
\label{R}\end{equation}
and
\begin{equation}
   \vec{y}={1 \over \sqrt{2}}
   \frac{1}{(A-1/2)(B-1/2)-C^2}
   \pmatrix{D(B-1/2)+C E \cr \ E(A-1/2)+C D}.
\label{Y}
\end{equation}
Equation~(\ref{eq.4.1}) describes the generic distribution function of a pure
two-mode squeezed coherent state. 

It is worth noting that the two-dimensional Hermite polynomial
$H_{n_1,n_2}^{\{R\}}(y_1,y_2)$ can be expressed in terms of usual Hermite
polynomials or generalized Laguerre polynomials 
depending on the structure of the matrix $R$~\cite{[38]}.

Now we show that the photon distribution of a pure two-mode 
squeezed coherent light obtained in terms of the 
modulus squared of a two-dimensional Hermite polynomial can be expressed 
linearly in terms of Hermite polynomials of four variables. For this, we use the 
expression for the photon distribution that was obtained in~\cite{[21]} and is 
given in our notation by 
\begin{equation}   
P(n_1,n_2)=P(0,0) \, \frac{H_{n_1,n_2,n_3,n_4}^{\{\tilde{R}\}}   
(y_1,y_2,y_3,y_4)}{n_1!\,n_2!}\,.
\label{w4}
\end{equation} 
Here the probability to have no photons is 
\begin{equation}  
P(0,0)=\left[\mbox {det}\left(\sigma+ \frac{1}{2} E_4\right)\right]^{-1/2}   
\exp \left[-\langle\vec{Q}\rangle \left(2\sigma+E_4\right)^{-1}\langle\vec{Q}\rangle \right], 
\end{equation} 
where the elements of the quadrature dispersion 4$\times $4 matrix $\sigma$
are given
by formula~(\ref{eq.sigma}). The matrix $\tilde{R}$ and the vector 
$\vec{y}=(y_1,y_2,y_3,y_4)$ are given by the expressions 
\begin{eqnarray}   
\tilde{R} & = & U^\dagger \, (E_4 - 2 \, \sigma)^{-1} \,    U^*\,,   
\nonumber \\
&&\\   \vec y & = & 2U^T(E_4-2\sigma)^{-1}\langle\vec{Q}\rangle \,,
\nonumber 
\end{eqnarray} 
with 
\begin{equation}    
U=\frac{1}{\sqrt{2}} \, \pmatrix{-iE_2 & iE_2 \cr E_2&E_2}. 
\end{equation} 
If we insert the dispersion matrix $\sigma$~~(\ref{eq.sigma}),
we arrive at the matrix
$\tilde{R}$ 
\begin{equation}   
\tilde{R} = \pmatrix{ r & 0 \cr 0 & r^\ast}, 
\end{equation} with the $2 \times 2 $ matrix $r$ 
\begin{equation}   
r = \frac{1}{ (A+1/2)(B+1/2)-C^2 }   
\pmatrix{(A-1/2)(B+1/2)-C^2 & -C \cr    -C & (A+1/2)(B-1/2)-C^2 } .
\end{equation} We note that the matrix $r$ is identical to the matrix $R$ in Eq.~(\ref{R}). 
In this case, the argument $\vec{y}$ of the four-dimensional Hermite polynomial may be 
split in
\begin{equation}   
\vec{y} = \pmatrix{ \vec{Y} \cr \vec{Y}^\ast }, 
\end{equation} 
where 
\begin{equation}   
\vec{Y}={1 \over \sqrt{2}}\,   \frac{1}{(A-1/2)(B-1/2)-C^2}   
\pmatrix{D(B-1/2)+C E \cr \ E(A-1/2)+C D}, 
\end{equation} 
which is the same vector as in Eq.~(\ref{Y}). 
Thus, the photon distribution function expressed in terms of modulus 
squared of Hermite polynomials of two variables may be also expressed linearly in 
terms of Hermite polynomials of four variables with equal pairs of indexes. 

\section{Photon Distribution Function for squeezed correlated states averaged
over one mode}
 In experiments, it is the photon number in the one mode that is usually
 measured. In this section we will average the two-mode photon distribution
 function over one mode and obtain the probability distribution function
 for having $n$ photon in other mode. For this purpose, first we write the
 density matrix in coordinate representation
\begin{eqnarray}
&&\rho(x_1,x_2,x_1',x_2')=|N|^2\exp\left(-A x_1^2+D x_1+2C x_1 x_2+E X_2
-A^*x_1^{'2}-B^*x_2^{'2} \right.\nonumber\\
&&\left.+2C^*x_1'x_2'+D^*x_1'+E^*x_2'\right)
\end{eqnarray}
Taking $x_2$ equal to $x_2'$ and integrate over $x_2'$ we obtain the density
matrix of one mode averaged over other mode
\begin{eqnarray}
&&\rho(x_1,x_1')=\sqrt{\frac{(A_1B_1-C_1^2)}{\pi B_1}}\exp\Big(-
\frac{B_1D_1^2+A_1E_1^2+2C_1E_1D_1}{2(A_1B_1-C_1^2)}+
\frac{E_1^2}{2B_1}\Big)
\nonumber\\
&&\times\exp\left(-\left(A_1-\frac{C^2}{2B_1}+\frac{1}{2}\right)x_1^2+
-\left(A^*_1-\frac{C^{*2}}{2B_1}+\frac{1}{2}\right)x_1^2 +2x_1x_1'
\frac{|C|^2}{2B_1}\right.\nonumber\\
&&\left.+x_1\left(D+\frac{CE_1}{B_1}\right)
+ x_1'\left(D^*+\frac{C^*E_1}{B_1}\right)\right).
\end{eqnarray}
In order to obtain averaged photon statistics we take the integral
\begin{equation}
P_n=\int\psi_n(x_1')\psi^*_n(x_1)\rho(x_1,x_1')d~x_1~dx_1',
\end{equation}
where $\psi_n(x_1,x_1')$ is wave function of Fock states and arrive at
\begin{equation}
P_n=\frac{P_o}{n!} H_{nn}^{\{R\}}(y_1,y_2).
\end{equation}
where
$$P_0=2\sqrt{\frac{A_1B_1-C_1^2}{B_1(|a|^2-b^2)}}\exp\Big(-
\frac{B_1D_1^2+A_1E_1^2+2C_1D_1}{2(A_1B_1-C_1^2)}+\frac{E_1^2}{2B_1}+
\frac{\mbox{Re}(a^*_1d^2)+b^2|d|^2}{2(|a_1|^2-b^2)}\Big),$$
$$a=A+\frac{1}{2}-\frac{C^2}{2B_1},\,b=\frac{|C|}{2B_1},\,
d=D+\frac{CE_1}{B_1},\,f=A-\frac{1}{2}-\frac{C^2}{2B_1^2}.$$
We see that averaged over one mode photon number distribution function
is expressed through Hermite polynomials of two variables with $R$ matrix
equal to
\begin{equation}
R=\frac{1}{|a_1|^2-b^2}\pmatrix{|a|^2-a^*-b^2&-b\cr-b&|a|^2-a-b^2}
\end{equation}
and arguments of Hermite polynomials equal to
\begin{equation}
y_1=y^*_2=\frac{df^*+d^*b}{\sqrt2(|a|^2-b^2)}.
\end{equation}
The photon number mean in the mode is determined by the quadrature 
dispersions and is equal to
\begin{eqnarray}
&&\langle n_1\rangle=\frac12\left(\sigma_{p_1,p_1}+\sigma_{q_1,q_2}-1\right)\nonumber\\
&&=
\frac{((1-2A_1)^2+4A_2^2)B_1+4(1-A_1)C_1^2-
8A_2C_1C_2+4A_1C_2^2}{8(A_1B_1-C_1^2)}. 
\end{eqnarray}
The photon number dispersion is determined by the formula  
\begin{equation}
\sigma_{n^2_1}=\frac12\left(\sigma_{q_1,q_1}^2+\sigma_{p_1,p_1}^2+
2\sigma_{q_1,p_1}^2-\frac12\right). 
\end{equation} 
It takes the following value for the squeezed correlated states of the photon modes 
\begin{eqnarray}
&&\sigma_{n^2_1}=
\frac{1}{(A_1B_1-C_1^2)^2}\left[-\frac14+\frac1{32}B_1^2+
\frac14(A_1B_2-C_1C_2)^2\right.\nonumber\\
&&+
\left.\frac12(A_1^2B_1+A_2(A_2B_1-2C_1C_2)+A_1(C_2^2-C_1^2))^2
\right] 
\end{eqnarray}
The Fano factor is the ratio of the photon dispersion and the photon number mean 
$$F_{Fano}=\frac{\sigma_{n^2_1}}{\langle n_1\rangle}$$
The Fano factor for the state under study reads
\begin{eqnarray}
&&F_{Fano}=\nonumber\\
&&\frac{-8(A_1B_1-C_1^2)+B_1^2+
8A_1B_2-C_1C_2+32(A_1^2B_1+A_2(A_2B_1-2C_1C_2)+
A_1(C_2^2-C_1^2))^2}{4(A_1B_1-C_1^2)
[((1-2A_1)^2+4A_2^2)B_1+
4(1-A_1)C_1^2-8A_2C_1C_2+4A_1C_2^2]}.\nonumber\\
&&\label{Fanofactor}
\end{eqnarray}
We see that the arguments of Hermite polynomial, matrix $R$, photon number mean, 
photon number dispersion and Fano factor are the
functions of coefficients in formulae~(\ref{eq.2.2}) ,determining the wave
function of squeezed correlated states.

\section{Entanglement in the two-mode squeezed correlated states}
In this section we discuss entanglement phenomena in the two-mode squeezed 
correlated states.
Entangled states are the states which are constructed as a superposition
of states each of which has the wave function expressed as a product of
wave functions depending on the different degrees of freedom.  We will employ 
two different simple measures of entanglement appropriate for the Gaussian states.
In \cite{[26]} the following measure of entanglement was suggested
\begin{equation}\label{ent_D}
E=\sigma_{q_1 q_2}^2+\sigma_{p_1 p_2}^2+\sigma_{q_1 p_2}^2
+\sigma_{p_1 q_2}^2.
\end{equation}
We employ the measure of entanglement (\ref{ent_D}) for evaluating the
correlations between the photon modes appearing in the two-mode 
squeezed correlated states.
Thus, one has the measure of entanglement of the modes in the form
\begin{eqnarray}
&&E_1=\frac{C_1^2+4(A_2C_1-A_1C_2)^2+4(B_2C_1-B_1C_2)^2}{16(A_1B_1-C_1^2)^2}\nonumber\\
&&+\frac{[A_2(B_2C_1-B_1C_2)+C_1(C_1^2+C_2^2)-A_1(B_1C_1+C_2(B_2+C_2))]^2
}{(A_1B_1-C_1^2)^2}.
\end{eqnarray}
In~\cite{[27]} another measure of entanglement was introduced as
the distance between the system density matrix and the tensor product
of the matrix partial traces over the subsystem degrees of freedom.
For the Gaussian states, the measure of entanglement reads
\begin{equation}\label{ent_S}
e_G=\frac{1}{4\sqrt{\mbox{det}\,\sigma(t)}}+\frac{1}{4\sqrt{\mbox{det}\,
\tilde{\sigma}}}-\frac{2}{\sqrt{\mbox{det}\,\Big(\sigma(t)
+\tilde{\sigma}\Big)}},
\end{equation}
where we use the block notation for the quadrature dispersion matrix
of the two-mode system
$$ \sigma(t)^{-1}=\left(\begin{array}{cc} \beta&\alpha\\ \alpha^T&\gamma
\end{array}\right),\qquad
\tilde{\sigma}=\left(\begin{array}{cc} \sigma_1 &0\\ 0&\sigma_2\end{array}\right)\,,$$
$$ \sigma_1^{-1}=\beta-\alpha \gamma^{-1}\alpha^T,\qquad \sigma_2^{-1}=
\gamma-\alpha^T \beta^{-1}\alpha.$$
Inserting the expressions for quadrature dispersions~(\ref{eq.sigma}) 
into~(\ref{ent_S}) one obtains the explicit formulas for the measure
of entanglement.
The expressions obtained for both measures of entanglement demonstrate
that the intermode interaction creates nonzero mode correlations
in the two-mode squeezed correlated states. 

\section{Symplectic tomogram of two-mode squeezed correlated states}
In this section, we apply the symplectic tomography scheme to consideration
of two-mode squeezed correlated states.
In the tomographic representation the quantum state is reconstructed
employing a probability distribution function called tomographic probability
(or marginal distribution). Such a representation was
introduced~\cite{[28]} in signal analysis. In~\cite{[29]} the
Wigner function of a quantum state was expressed in terms of measurable
experimantally tomographic probability. In~\cite{[31]} this idea was
realised experimentally and the method of measuring quantum state of
photon was called the optical tomography  method. The optical tomography
was used in~\cite{[32]} to measure squeezed mixed states of photon.
In our work we discuss the case of tomographic probability distribution for
two-mode system. We use symplectic tomography scheme~\cite{[35],[36]},
which is the generalization of optical tomography scheme, and take into
account the results of \cite{[39],[40]}.The tomogram of the quantum state of
photons is nonnegative probability distribution of the two quadratures
$X_1$ and $X_2$ (tomographic quadratures).
It is measured in the reference frame in the phase space of the quadrature
components of the photons which is labeled by four real parameters. The
quadrature $X_1$ can be interpreted as the eigenvalue of rotated and scaled
quadrature operator of the first photon mode. The quadrature $X_2$ can be
interpreted as the eigenvalue of rotated and scaled quadrature operator of
the other photon mode. The two operators read
\begin{equation}\label{eq.st14}
\hat X_1=\mu_1q_1+\nu_1p_1\,,\qquad
\hat X_2=\mu_2q_2+\nu_2p_2\,.
\end{equation}
For $\mu=\cos \theta$, $\nu=\sin\theta$ (in the one-mode case),
the tomogram $w(X_1,\cos\theta,\sin\theta)$ is the measurable probability
distribution of the optical tomography procedure~\cite{[29],[31]},
in which the measurable observable $\hat X_1$ takes the form of rotated
photon quadrature operator
\begin{equation}\label{eq.st15}\hat X_{1\theta}=q_1~\cos\theta +
p_1~\sin\theta.
\end{equation}
Such photon quadrature can be measured by means of the homodyne
photon detection scheme~\cite{[31]}. The density matrix of both
photon modes determines the Wigner function of the system
\begin{equation}\label{eq.sp6}
W_{12}(q_1,q_2,p_1,p_2)= \int \rho_{12}\left(q_1+\frac{u_1}{2},q_2+
\frac{u_2}{2},
q_1-\frac{u_1}{2},q_2-\frac{u_2}{2}\right)
e^{-i p_1 u_1-i p_2 u_2}\,d u_1\,du_2,
\end{equation}
where the density matrix $\rho(x_1,x_2,x_1',
x_2')$ is considered
in the position representation. The coordinates $x_1,\,x_1'$ are used for the one photon mode and the
coordinates $x_2,\,x_2'$ are used for other photon mode. The tomogram of the
two-mode system is given in terms of the
Wigner function by the relation
\begin{eqnarray}\label{eq.sp12}
&&w_{12}(X_1,X_2,\mu_1,\mu_2,\nu_1,\nu_2) =\frac{1}{4\pi^2} \int
W_{12}(q_1,q_2,p_1,p_2)\nonumber\\
&&\times\delta(X_1-\mu_1 q_1 -\nu_1 p_1)\delta(X_2-\mu_2 q_2-\nu_2 p_2)\,
d q_1\, d q_2\, d p_1\, d p_2.
\end{eqnarray}
Here we use the following notation. The parameters $\mu_1$ and
$\nu_1$ describe the rotated and scaled reference frame
in the phase space of the photon quadratures of first mode.
The parameters $\mu_2$ and $\nu_2$ describe the rotated and scaled
reference frame in the phase space of the photon quadratures of second mode.
The Wigner function of the two-mode system is determined by the tomogram due
to relation, which is inverse of (\ref{eq.sp12})
\begin{eqnarray}\label{eq.sp11}
&&W_{12}(q_1,q_2,p_1,p_2) =
\frac{1}{4\pi^2}\int w_{12}(X_1,X_2,\mu_1,\mu_2,\nu_1,\nu_2)\nonumber\\
&&\times\exp\left[ i(X_1-\mu_1 q_1 -\nu_1 p_1+X_2-\mu_2 q_2-\nu_2 p_2)\right]
d X_1\,d X_2\, d\mu_1\, d\mu_2\, d\nu_1\,d\nu_2.\nonumber\\
\end{eqnarray}
One can show that the tomogram of the photon subsystem is related
to the tomogram of complete system by the relation
\begin{equation}\label{eq.sp14}
w_{1}(X_1,\mu_1,\nu_1) = \int w_{12}(X_1,X_2,\mu_1,\mu_2,\nu_1,
\nu_2)\,d X_2\,,
\end{equation}
The tomogram of the two-mode system for Gaussian density
matrix has the form of the standard two-dimensional Gaussian distribution
\begin{equation}\label{eq.sg13}
 w_G(X_1,X_2,\mu_1,\mu_2,\nu_1,\nu_2)=
\frac{1}{2\pi\sqrt{\mbox{det}\,\sigma_{\bf X}}}
\exp\left[-{1\over 2}({\bf X}-\langle{\bf X}\rangle)
\sigma_{\bf X}^{-1} ({\bf X}-\langle{\bf X}\rangle)\right],
\end{equation}
determined by means and dispersions of the random variables
${\bf X}=(X_1,X_2)$.
In (\ref{eq.sg13}) one has
\begin{equation}\label{eq.sg14}
\langle{\bf X}\rangle= \pmatrix{\mu_1\langle q_1\rangle+\nu_1\langle p_1
\rangle \cr
\mu_2\langle q_2\rangle+\nu_2\langle p_2\rangle},
\end{equation}
where $\langle q_1\rangle,\,\langle p_1\rangle$ are quadrature means of the
first photon mode and $\langle q_2\rangle,\,\langle p_2\rangle$ are
quadrature means of other photon mode.
Inserting the expressions for quadrature means~(\ref{means})
into~(\ref{eq.sg14}) we obtain 
\begin{eqnarray} \label{eq.sg14a}
\langle X \rangle_1 &=& \frac{(B_1D_1+C_1E_1)
(\mu_1-2\nu_1A_2)}{2(A_1B_1-C_1^2)} + 
\nu_1 \left( D_2 +\frac{C_2(E_1A_1+D_1C_1)}{2(A_1B_1-C_1^2)}\right), \nonumber\\
\langle X \rangle_2 &=& \frac{(A_1E_1+C_1D_1)(\mu_2-2\nu_2B_2)}{2(A_1B_1-
C_1^2)} + \nu_2 \left( E_2 +\frac{C_2(D_1B_1+E_1C_1)}{2(A_1B_1-C_1^2)}\right) 
\end{eqnarray}
The matrix elements of the
symmetric dispersion matrix
\begin{equation}\label{eq.sg15}
\sigma_{\bf X}=\pmatrix{\sigma_{X_1^2}&\sigma_{X_1 X_2}\cr
\sigma_{X_1 X_2}&\sigma_{X_2^2}}
\end{equation}
are variances
\begin{equation}\label{eq.sg16}
\sigma_{X_1^2}=\mu_1^2\sigma_{q_1^2}+
\nu_1^2\sigma_{p_1^2}+2\mu_1\nu_1\sigma_{q_1 p_1}\,,
\qquad
\sigma_{X_2^2}=\mu_2^2\sigma_{q_2^2}+\nu_2^2\sigma_{p_2^2}+
2\mu_2\nu_2\sigma_{q_2 p_2}
\end{equation}
and covariance
\begin{equation}\label{eq.sg18}
\sigma_{X_1 X_2}=\mu_1\mu_2\sigma_{q_1 q_2}+
\nu_1\nu_2\sigma_{p_1 p_2}+\mu_1\nu_2\sigma_{q_1 p_2}+
\mu_2\nu_1\sigma_{q_2 p_1}
\end{equation}
of the photon and phonon tomographic quadratures.
Inserting the expressions for quadrature dispersions~(\ref{eq.sigma})
into~(\ref{eq.sg16},\ref{eq.sg18}) we obtain
\begin{eqnarray}\label{eq.sg14b}
&&\sigma_{X_1X_2}= \frac{B_1(\mu_1\mu_2-4\nu_1\nu_2A_2C_2+
2\mu_1\nu_2C_2)}{4(A_1B_1-C_1^2)} - 
\frac{C_1(\mu_1\nu_2B_2-\nu_1\nu_2C_2^2)}{2(A_1B_1-C_1^2)}\nonumber\\
&&- \nu_1\nu_2\left(C_1+\frac{A_1C_2^2}{A_1B_1-C_1^2}\right)\cr 
\sigma_{X_1^2} &=& \frac{B_1(\mu_1-2\nu_1A_2)^2}{4(A_1B_1-C_1^2)} + 
\frac{\nu_1C_1C_2(\mu_1-2\nu_1 A_2)}{A_1B_1-C_1^2} +\nu_1^2 A_1
 \left(1 +\frac{C_2^2}{A_1B_1-C_1^2}\right) \cr 
\sigma_{X_2^2} &=& \frac{A_1(\mu_2-2\nu_2B_2)^2}{4(A_1B_1-C_1^2)}+
\frac{\nu_2C_1C_2(\mu_2-2\nu_2B_2)}{A_1B_1-C_1^2}+\nu_2^2 B_1\left(1 + 
\frac{C_2^2}{A_1B_1-C_1^2}\right)\nonumber\\
&&\label{XX}
\end{eqnarray}
Inserting~(\ref{eq.sg14a},\ref{eq.sg14b}) into~(\ref{eq.sg13}) the tomogram
of two-mode squeezed correlated states can be obtained in explicit form.

\section{Two-mode even and odd coherent states}
In this section we will consider the two-mode even and odd coherent states. 
We define the two-mode even and odd coherent states as the simplest
superposition of two-mode coherent states 
\begin{equation}
|\alpha_1,\alpha_2\rangle_{\pm} =
N_{\pm}\left(|\alpha_1,\alpha_2\rangle\pm|-\alpha_1,-
\alpha_2\rangle\right), 
\end{equation}
here $\alpha_1$ and $\alpha_2$ are complex numbers. For even coherent
states one has to take plus in superposition, and for odd coherent
states one has to take minus.
The normalization constants for two-mode even and odd coherent states are 
\begin{equation}
N_+=\frac{\exp\left(\frac{|\alpha_1|^2}2+
\frac{|\alpha_2|^2}2\right)}{2\sqrt{\cosh\left(|\alpha_1|^2
+|\alpha_2|^2\right)}};
\quad N_-=\frac{\exp\left(\frac{|\alpha_1|^2}2+
\frac{|\alpha_2|^2}2\right)}{2\sqrt{\sinh\left(|\alpha_1|^2+
|\alpha_2|^2\right)}};
\end{equation}
We can write the wave functions of even and odd coherent states
in following explicit forms 
\begin{eqnarray}
|\alpha_1,\alpha_2\rangle_+&=&
\frac{N_{+}}{\sqrt\pi}\exp\left(-\frac{q_1^2}2-\frac{q_2^2}2
-\frac{|\alpha_1|^2}2-\frac{|\alpha_2|^2}2-\frac{\alpha_1^2}2-
\frac{\alpha_2^2}2\right)
\cosh\left(\sqrt{2}\,x_1\alpha_1+\sqrt{2}\,x_2\alpha_2\right)\nonumber\\
|\alpha_1,\alpha_2\rangle_-&=&
\frac{N_{-}}{\sqrt\pi}\exp\left(-\frac{q_1^2}2-\frac{q_2^2}2
-\frac{|\alpha_1|^2}2-\frac{|\alpha_2|^2}2-\frac{\alpha_1^2}2-
\frac{\alpha_2^2}2\right)
\sinh\left(\sqrt{2}\,x_1\alpha_1+\sqrt{2}\,x_2\alpha_2\right).\nonumber\\
&&\label{tom}
\end{eqnarray}
The Wigner functions for even and odd coherent states are
\begin{eqnarray}
&&W_{\alpha_1,\alpha_2\pm}=4|N_\pm|^2
\exp\left(-q_1^2-q_2^2-p_1^2-p_2^2\right)\nonumber\\
&&\times\left\{\exp\left(-2|\alpha_1|^2-2|\alpha_2|^2\right)
\cosh\left[2\sqrt2\,\left(\mbox{Re}[\alpha_1]q_1+\mbox{Re}[\alpha_2]q_2
+\mbox{Im}[\alpha_1]p_1+\mbox{Im}[\alpha_2]p_2\right)\right]\right. \nonumber\\
&&\pm
\left.\cos\left[2\sqrt2\,\left(\mbox{Im}[\alpha_1]q_1+\mbox{Im}[\alpha_2]q_2
-\mbox{Re}[\alpha_1]p_1-\mbox{Re}[\alpha_2]p_2\right)\right]\right\}.
\end{eqnarray}
One can see that the Wigner functions for even and odd coherent states
are a sum of four gaussians.
The probabilities of finding $n_1$ photons in the first mode and $n_2$
photons in the second mode are
\begin{eqnarray}
P_+(n_1, n_2)&=&\frac{|\alpha_1|^{2n_1}|
\alpha_2|^{2n_2}}{n_1!n_2!
\cosh\left(|\alpha_1|^2+|\alpha_2|^2\right)},\quad n_1+n_2=2k,\\
P_-(n_1,n_2)&=&\frac{|\alpha_1|^{2n_1}|
\alpha_2|^{2n_2}}{n_1!n_2!\sinh(|\alpha_1|^2+|\alpha_2|^2)},
\quad n_1+n_2=2k+1\nonumber
\end{eqnarray}
We see that in the present case of two-mode even and odd coherent states we
cannot factorize their photon distribution functions due to the presence of
the nonfactorizable $\cosh(|\alpha_1|^2+|\alpha_2|^2)$ and
$\sinh(|\alpha_1|^2+|\alpha_2|^2)$. This fact implies the phenomenon of
statistical dependences of different modes of these states one on each other. 
These probabilities are equal to zero, if the sum of $n_1$ and $n_2$ is an
odd number for even states, or if this sum is an even number in the case of
odd states.
After the averaging over one mode we obtain the probability to have $n$ photon
in other mode in the form
\begin{eqnarray}
\tilde{P}_+(n_1)&=&\sum\limits_{n_2=0}^\infty
P_+(n_1, n_2)=\frac{|\alpha_1|^{2n_1}e^{|\alpha_2|^2}}{n_1!\cosh
\left(|\alpha_1|^2+|\alpha_2|^2\right)}\\
\tilde{P}_-(n_1)&=&\sum\limits_{n_2=0}^\infty P_+(n_1, n_2)=
\frac{|\alpha_1|^{2n_1}e^{|\alpha_2|^2}}{n_1!\sinh
\left(|\alpha_1|^2+|\alpha_2|^2\right)}\nonumber
\end{eqnarray}
The dispersions of quadrature components in the two-mode even coherent
states are
$$\sigma_{q_1^2}=|\alpha_1|^2\tanh(|\alpha_1|^2+|\alpha_2|^2)+
\mbox{Re}[\alpha_1^2]+\frac12,$$
$$\sigma_{q_2^2}=|\alpha_2|^2\tanh(|\alpha_1|^2+|\alpha_2|^2)+
\mbox{Re}[\alpha_2^2]+\frac12,$$
$$\sigma_{q_1,q_2}=\left(\tanh(|\alpha_1|^2+|\alpha_2|^2)+1\right)
\mbox{Re}[\alpha_1\alpha_2],$$
$$\sigma_{p_1^2}=|\alpha_1|^2
\tanh(|\alpha_1|^2+|\alpha_2|^2)-\mbox{Re}[\alpha_1^2]+\frac12,$$
$$\sigma_{p_2^2}=|\alpha_2|^2
\tanh(|\alpha_1|^2+|\alpha_2|^2)-\mbox{Re}[\alpha_2^2]+\frac12,$$
$$\sigma_{p_1,p_2}=\left(\tanh(|\alpha_1|^2+|\alpha_2|^2)-1\right)
\mbox{Re}[\alpha_1\alpha_2],$$
$$\sigma_{q_1,p_1}=\mbox{Im}[\alpha_1^2],$$
$$\sigma_{q_2,p_2}=\mbox{Im}[\alpha_2^2],$$
$$\sigma_{q_1,p_2}=\sigma_{q_2,p_1}=\mbox{Im}[\alpha_1\alpha_2].$$
The dispersions of quadrature components in the two-mode odd coherent
states are
$$\sigma_{q_1^2}=|\alpha_1|^2\coth(|\alpha_1|^2+|\alpha_2|^2)+
\mbox{Re}[\alpha_1^2]+\frac12,$$
$$\sigma_{q_2^2}=|\alpha_2|^2\coth(|\alpha_1|^2+|\alpha_2|^2)+
\mbox{Re}[\alpha_2^2]+\frac12,$$
$$\sigma_{q_1,q_2}=\left(\coth(|\alpha_1|^2+|\alpha_2|^2)+1\right)
\mbox{Re}[\alpha_1\alpha_2],$$
$$\sigma_{p_1^2}=|\alpha_1|^2
\coth(|\alpha_1|^2+|\alpha_2|^2)-\mbox{Re}[\alpha_1^2]+\frac12,$$
$$\sigma_{p_2^2}=|\alpha_2|^2
\coth(|\alpha_1|^2+|\alpha_2|^2)-\mbox{Re}[\alpha_2^2]+\frac12,$$
$$\sigma_{p_1,p_2}=\left(\coth(|\alpha_1|^2+|\alpha_2|^2)-1\right)
\mbox{Re}[\alpha_1\alpha_2],$$
$$\sigma_{q_1,p_1}=\mbox{Im}[\alpha_1^2],$$
$$\sigma_{q_2,p_2}=\mbox{Im}[\alpha_2^2],$$
$$\sigma_{q_1,p_2}=\sigma_{q_2,p_1}=\mbox{Im}[\alpha_1\alpha_2].$$
We see that the formulae for dispersions of quadratures in odd
coherent states can be obtain from formulae for dispersions of quadratures
in even coherent states by changing the function $\tanh$ by the function
$\coth$.
The photon number means for even and coherent state are  
$$\langle n_1\rangle_+=|\alpha_1|^2\tanh(|\alpha_1|^2+|\alpha_2|^2).$$
$$\langle n_1\rangle_-=|\alpha_1|^2\coth(|\alpha_1|^2+|\alpha_2|^2).$$
The photon number dispersions in the two-mode odd and even coherent states
are 
$$\sigma_{n_1^2+}
=|\alpha_1|^4\mbox{sech}^2(|\alpha_2|^2+|\alpha_1|^2)+
|\alpha_1|^2\tanh(|\alpha_1|^2+|\alpha_2|^2),$$
$$\sigma_{n_1^2-}
=|\alpha_1|^2\coth(|\alpha_1|^2+|\alpha_2|^2)
  - |\alpha_1|^4\mbox{cosech}^2(|\alpha_2|^2+|\alpha_1|^2).$$
We can calculate the Fano factor in even and odd coherent states
$$F_+=1+\frac{|\alpha_1|^2\mbox{sech}^2(|\alpha_1|^2+|\alpha_2|^2)}
{\tanh(|\alpha_1|^2+|\alpha_2|^2)},$$
$$F_-=1-\frac{|\alpha_1|^2\mbox{cosech}^2(|\alpha_1|^2+|\alpha_2|^2)}
{\coth(|\alpha_1|^2+|\alpha_2|^2)}.$$
%
 %  \begin{figure}
  % \begin{center}
   %\begin{tabular}{cc}
   %\includegraphics[height=6cm]{Fp_cat.eps}
   %\includegraphics[height=6cm]{Fm_cat.eps}
  % \end{tabular}
  % \end{center}
  % \caption[Fano factor]
   %{\label{fig:F} The Fano factor for even (left) and odd (right) coherent state with $\alpha_1$ and $\alpha_2$ %supposed %to be real.}
  % \end{figure}
%
The Fano factor for even coherent states is more
then unity for all values of $\alpha_1,\quad \alpha_2$,
so the photon distribution
function in even coherent states is always super-Poissonian. The Fano
factor for odd coherent states is less then unity for all
values of $\alpha_1,\quad \alpha_2$, so the photon distribution
function in odd coherent states is always sub-Poissonian. 
The measures of entanglement in even and odd coherent states exhibited the
appearance of correlations between the modes and are of the form 
$$E_+=2\mbox{Re}^2[\alpha_1,\alpha_2]\mbox{sech}^2(|\alpha_1|^2+|\alpha_2|^2) +
2\mbox{Im}^2[\alpha_1,\alpha_2],$$
$$E_-=2\mbox{Re}^2[\alpha_1,\alpha_2]\mbox{cosech}^2(|\alpha_1|^2+|\alpha_2|^2) +
2\mbox{Im}^2[\alpha_1,\alpha_2],$$
 %  \begin{figure}
   %\begin{center}
   %\begin{tabular}{cc}
   %\includegraphics[height=6cm]{Ep_cat.eps}
   %\includegraphics[height=6cm]{Em_cat.eps}
   %\end{tabular}
   %\end{center}
   %\caption[Entanglement]
   %{\label{fig:Ent}Entanglement for odd (left) and even (right) coherent state.  Here $\alpha_1$ and $\alpha_2$ are %taken to %be real}
  % \end{figure}
Let us apply symplectic tomography scheme to the even and odd coherent
states and obtain the tomogram of two-mode even and odd coherent
states in explicit form  
$$\omega_{0\pm}=8|N_\pm|^2\exp\left(-|\alpha_1|^2-|\alpha_2|^2\right)\frac{\pi\mu_1\mu_2}{\sqrt{\mu_1^2+\nu_1^2}\sqrt{\mu_2^2+\nu_2^2}}\exp\left(-\frac{x_1^2}{\mu_1^2}-\frac{x_2^2}{\mu_2^2}\right)\times$$
$$\times\exp\left[\frac1{\mu_1^2+\nu_1^2}\left(\frac{\nu_1^2x_1^2}{\mu_1^2}+2\nu_1^2\mbox{Re}^2\alpha_1+2\mu_1^2\mbox{Im}^2\alpha_1-4\mu_1\nu_1\mbox{Re}\alpha_1\mbox{Im}\alpha_1\right)+\right.$$
$$\left.+\frac1{\mu_2^2+\nu_2^2}\left(\frac{\nu_2^2x_2^2}{\mu_2^2}+2\nu_2^2\mbox{Re}^2\alpha_2+2\mu_2^2\mbox{Im}^2\alpha_2-4\mu_2\nu_2\mbox{Re}\alpha_2\mbox{Im}\alpha_2\right)\right]\times$$
$$\times\cosh\left[2\sqrt2\left(\frac{x_1}{\mu_1}\mbox{Re}\alpha_1+\frac{x_2}{\mu_2}\mbox{Re}\alpha_2+
\frac{x_1\nu_1}{\mu_1}\frac1{\mu_1^2+\nu_1^2}\left(\mu_1\mbox{Im}\alpha_1-\nu_1\mbox{Re}\alpha_1\right)\right.\right.$$
$$+
\left.\left.\frac{x_2\nu_2}{\mu_2}\frac1{\mu_2^2+\nu_2^2}\left(\mu_2\mbox{Im}\alpha_2-\nu_2\mbox{Re}\alpha_2\right)
\right)\right]\pm$$
$$\pm8|N_\pm|^2\frac{\pi\mu_1\mu_2}{\sqrt{\mu_1^2+\nu_1^2}\sqrt{\mu_2^2+\nu_2^2}}\exp\left(-\frac{x_1^2}{\mu_1^2}-\frac{x_2^2}{\mu_2^2}\right)\times$$
$$\times\exp\left[\frac1{\mu_1^2+\nu_1^2}\left(\frac{\nu_1^2x_1^2}{\mu_1^2}-2\mu_1^2\mbox{Re}^2\alpha_1-2\nu_1^2\mbox{Im}^2\alpha_1-4\mu_1\nu_1\mbox{Re}\alpha_1\mbox{Im}\alpha_1\right)+\right.$$
$$\left.+\frac1{\mu_2^2+\nu_2^2}\left(\frac{\nu_2^2x_2^2}{\mu_2^2}-2\mu_2^2\mbox{Re}^2\alpha_2-2\nu_2^2\mbox{Im}^2\alpha_2-4\mu_2\nu_2\mbox{Re}\alpha_2\mbox{Im}\alpha_2\right)\right]\times$$
$$\times\cos\left[2\sqrt2\left(\frac{x_1}{\mu_1}\mbox{Im}\alpha_1+\frac{x_2}{\mu_2}\mbox{Im}\alpha_2-
\frac{x_1\nu_1}{\mu_1}\frac1{\mu_1^2+\nu_1^2}\left(\mu_1\mbox{Re}\alpha_1+\nu_1\mbox{Im}\alpha_1\right)\right.\right.$$
$$-
\left.\left.\frac{x_2\nu_2}{\mu_2}\frac1{\mu_2^2+\nu_2^2}\left(\mu_2\mbox{Re}\alpha_2+\nu_2\mbox{Im}\alpha_2\right)
\right)
\right]$$
These tomograms are the images of the nonclassical even and odd coherent
states in the probability representation of quantum mechanics. 
We can use tomograms for describing even and odd coherent states
instead of using Wigner functions or wave function of the states.

\section{Conclusions}
We discuss photon-number probability distribution function for
two-mode even and odd coherent states and two-mode squeezed
correlated states and after averaging over one mode we obtain
the probability of having $n$ photons in other mode
for the states under study. We calculated means, dispersions of quadrature
components and of photon numbers in the modes, Fano factors and
tomograms within the framework of symplectic tomography scheme for
two-mode even and odd coherent states and two-mode squeezed
correlated state. We saw that for the two-mode even coherent
states the photon statistics is always super-poissonian and for the
two-mode odd coherent states it is always sub-poissonian. We evaluated
the measure of entanglement employing two different methods. The
nonzero measure of entanglement shows the statistical dependence
between the modes in the states under study.

\section*{Acknowledgments}

This study was supported by the Russian Foundation for Basic Research
under Project Nos.~01-02-17745 and 03-02-16408.

\end{document}